\documentclass[10pt,letterpaper]{article}
\usepackage{opex3}
\usepackage{color}
\usepackage[colorlinks=true,urlcolor=blue,linkcolor=black,citecolor=black,breaklinks=true,bookmarks=false]{hyperref}
\usepackage[euler]{textgreek}
\usepackage{cite}

\begin{document}

\title{Raman-assisted coherent, mid-infrared frequency combs in silicon microresonators}

\author{Austin G. Griffith$^1$, Mengjie Yu$^{2,3}$, Yoshitomo Okawachi$^3$, \\Jaime Cardenas$^4$, Aseema Mohanty$^{2,4}$, Alexander L. Gaeta$^3$, \\and Michal Lipson$^{4}$}

\address{$^1$School of Applied \& Engineering Physics, Cornell University, Ithaca, NY 14853, USA\\
$^2$School of Electrical \& Computer Engineering, Cornell University, Ithaca, NY 14853, USA\\
$^3$Department of Applied Mathematics and Applied Physics, Columbia University, New York, NY 10027, USA\\
$^4$Department of Electrical Engineering, Columbia University, New York, NY 10027, USA\\}

\email{$^*$ml3745@columbia.edu} 



\begin{abstract}
We demonstrate the first low-noise mid-IR frequency comb source using a silicon microresonator.  Our observation of strong Raman scattering lines in the generated comb suggests that Raman and four-wave mixing interactions play a role in assisting the transition to the low-noise state. In addition, we characterize, the intracavity comb generation dynamics using an integrated PIN diode, which takes advantage of the inherent three-photon absorption process in silicon.
\end{abstract}



There is significant interest in mid-infrared (mid-IR) frequency comb technology for applications in high-resolution spectroscopy and metrology \cite{Schliesser, Keilmann,Adler}. Using a four-wave mixing (FWM) process based on the third-order nonlinearity $\chi^{(3)}$ in a microresonator, one can generate a broadband frequency comb using parametric oscillation \cite{Kippenberg}. This has been realized in a number of different material platforms in the near-infrared \cite{Savchenkov11,Saha1um}, at telecommunication wavelengths \cite{PWang,Saha,Jung,Hausmann,Herr,Papp,Yi} and in the mid-IR \cite{CWang,Griffith,Luke,Savchenkov,Lecaplain}, offering promise for a compact, chip scale comb source in the mid-IR. However, the generated microresonator-based comb is not always low-noise or phase-coherent. For example, if the comb sidebands are generated multiple mode spacings from the pump laser, the resulting comb will not initially be coherent. Here, each of the sidebands can generate its own `mini-comb' with a different carrier-envelope offset frequency \cite{Herr12}. When these mini-combs merge, the resulting frequency comb lines will be broadened by the existence of multiple comb lines per microresonator resonance. This can be experimentally detected in the time domain or by the presence of radio frequency (RF) amplitude noise in the frequency comb arising from the beating of these overlapped mini-combs \cite{Saha,Herr12}. A high-noise microresonator comb can, in principle, transition to a low-noise state, where the overlapped mini-combs lock together, and the comb spacing across the comb equalizes. After this low-noise transition, the emergence of optical pulse trains and temporal solitons have been observed \cite{Saha,Herr,Yi}. Here, we demonstrate the first low-noise coherent mid-IR frequency comb source using a silicon microresonator. We observe strong comb lines separated by the Raman shift in silicon, indicating that interactions between the Raman effect and FWM results in phase locking of the generated comb. In addition, we introduce a novel technique for characterizing the intracavity comb generation dynamics using an integrated PIN diode, which utilizes the free-carriers (FC's) generated through the inherent three-photon absorption process in silicon.    

\begin{figure}[tp]
\centering\includegraphics[width=12cm]{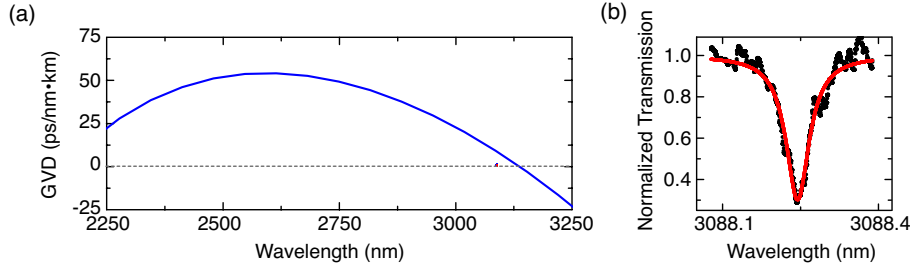}
\caption{(a) The simulated dispersion parameter of the etchless waveguide. The group-velocity dispersion (GVD) is anomalous (\emph{i.e.}, >0) past 3.1 \textmu m. (b) We characterize an overcoupled microresonator at 3.1 \textmu m with a loaded quality factor of 60,000. The resonance is confirmed to be overcoupled by using the PIN diode to inject carriers into the resonance, and observing an increase in the resonance extinction. This equates to an intrinsic quality factor of 250,000.}
\label{fig:fig1}
\end{figure}

For comb generation in the mid-IR, we use a silicon microresonator with an integrated PIN diode. We fabricate our devices using an ÔetchlessÕ process that uses thermal oxidation instead of reactive-ion etching to form the waveguide core \cite{Griffith,Griffith12}. The etchless waveguide is dispersion engineered to allow for anomalous group-velocity dispersion (GVD) at the pump wavelength enabling comb generation [Fig. \ref{fig:fig1}(a)]. We characterize an overcoupled resonance of a silicon microresonator at 3.1 \textmu m wavelength and measure a loaded quality factor of 60,000, corresponding to an intrinsic quality factor of 250,000 [Fig. \ref{fig:fig1}(b)]. Silicon suffers from three-photon absorption (3PA) in the 2.2 to 3.3 \textmu m wavelength regime, and the generated photocarriers can cause significant FC absorption (FCA) for long carrier lifetimes. To mitigate this, the silicon microresonators are embedded in an integrated PIN diode to enable extraction of the generated free carriers \cite{Griffith,Rong05,TurnerFoster}. When a reverse-bias voltage is applied to the PIN junction, carriers are swept out of the diode depletion region. The electrical contacts for the PIN diode are spaced 4.4 \textmu m apart, with the etchless waveguide in the center. 

\begin{figure}[tp]
\centering\includegraphics[width=12cm]{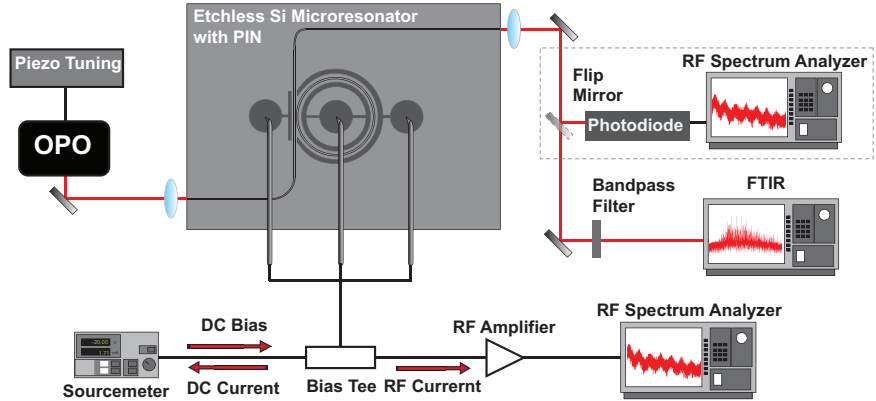}
\caption{Experimental setup for generation and characterization of mid-IR frequency comb in silicon microresonators. We pump a silicon microresonator using a cw optical parametric oscillator (OPO). The output is collected using an FTIR. We monitor the RF noise using a conventional photodiode and a PIN diode.}
\label{fig:fig2}
\end{figure}

The experimental setup for measuring the optical and RF spectra of the generated comb is shown in Fig. \ref{fig:fig2}. For comb generation, we pump the silicon microresonator with a single-frequency optical parametric oscillator (Argos Model 2400). The optical spectrum is recorded using a Fourier transform infrared spectrometer (FTIR), along with a series of 500-nm bandwidth bandpass filters to resolve different sections of the generated frequency comb. In addition, the output from microresonator is measured with a commercial InGaAs photodetector and then sent to an RF spectrum analyzer. The detector bandwidth is 10 MHz. Unlike in the near-infrared, photodetectors in the mid-IR have a significantly narrower bandwidth, restricting the RF frequency range over which the comb dynamics can be characterized. In our microresonator structure, we can circumvent this limitation by measuring the photocurrent from the generated free carriers. By applying a reverse-bias voltage at the PIN junction, 3PA-induced free carriers are extracted and a photocurrent is generated. The RF component of the 3PA-induced current is extracted using a bias-tee and sent to a second RF spectrum analyzer, which allows characterization of the intracavity power. Based on the 3PA, the FC density is proportional to the cube of the optical intensity within the cavity. Therefore, fluctuations in intracavity power will be reflected in the noise of 3PA-induced current. The detection bandwidth is largely determined by the FC lifetime, which can be controlled with the reverse-bias voltage. In addition, the wavelength response covers 2.2--3.3 \textmu m due to 3PA in silicon.

\begin{figure}[tp]
\centering\includegraphics[width=10cm]{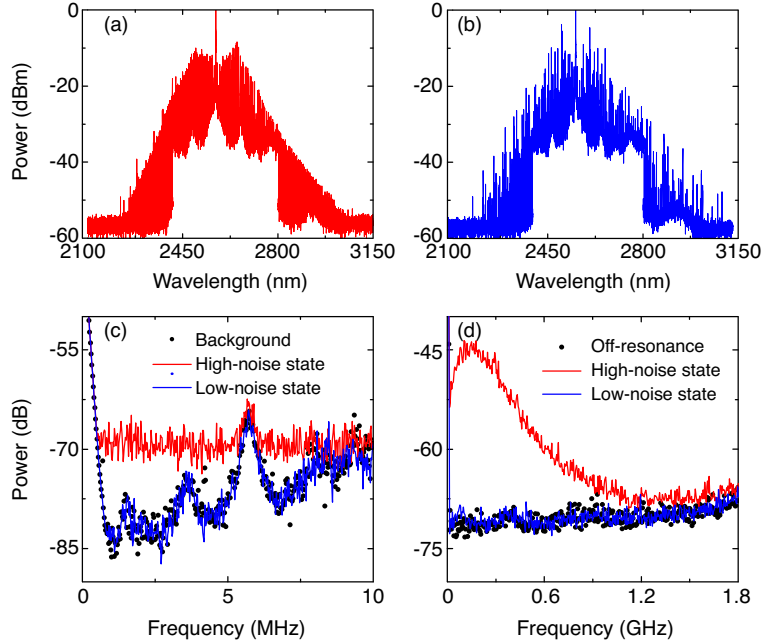}
\caption{(a) High-noise and (b) low-noise combs generated pumping at 2.6 \textmu m. We observe a reduction in RF noise in both (c) photodetector and (d) PIN detector. The photodiode measurement is limited by the 10 MHz bandwidth of the photodetector.}
\label{fig:fig3}
\end{figure}

First, we investigate the RF characteristics of the generated comb using both the commercial photodiode and the PIN diode. We pump a silicon microresonator at 2.6 \textmu m with 180 mW in the bus waveguide at a reverse-bias voltage of -12 V. We observe two different comb states [Fig. \ref{fig:fig3}(a) \& (b)] which correspond to high- and low-noise states, respectively. The RF noise transition is observed simultaneously in both PIN-based and photodiode-based RF measurements [Fig. \ref{fig:fig3}(c) \& (d)]. In contrast, the RF signal from the PIN diode in Fig. \ref{fig:fig3}(d) is strong enough to be measured without any amplification up to 1.8 GHz, due to the large 3PA coefficient in silicon \cite{Pearl}. The correspondence of both measurements, albeit at different frequency scales, confirms that the PIN-based RF measurement is indeed measuring the actual RF state of the silicon microresonator comb. The discrepancy in RF features observed in the two different measurements can be attributed to the fact that the PIN-based photocurrent response is nonlinear since it relies on 3PA, and the wavelength range of the 3PA response is different from that of the photodiode response (1.2--2.6 \textmu m). Overall, this PIN-based measurement allows for direct monitoring of the processes occurring within the resonator and provides another means of characterizing the noise and phase-locking properties of the comb. 

\begin{figure}[tp]
\centering\includegraphics[width=13cm]{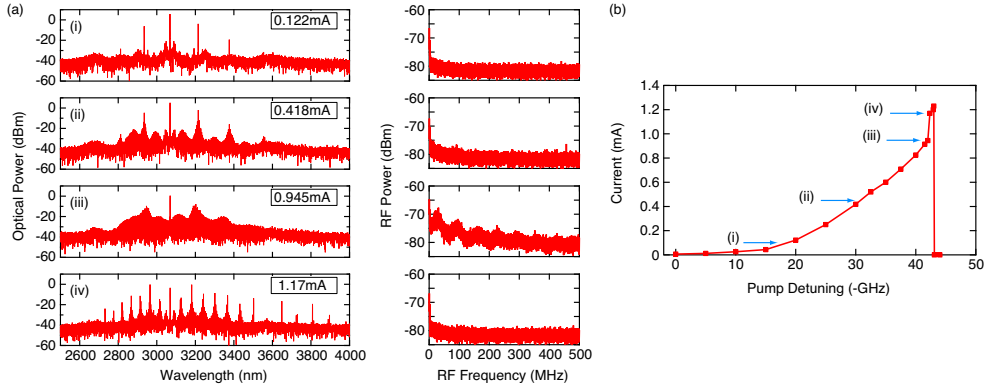}
\caption{Comb generation dynamics in a silicon microresonator. (a) Optical and RF spectra for generated comb as pump is red-detuned. (b) Measured DC component of the FC induced photocurrent as a function of pump detuning. We observe an abrupt increase in current as the comb transitions to a low-noise state.
}
\label{fig:fig4}
\end{figure}

We investigate the comb generation dynamics in the silicon microresonator for the case in which the pump wavelength is set to 3.07 \textmu m and the reverse-bias voltage on the PIN structure is set to -12 V. As the pump wavelength is tuned into a cavity resonance, we characterize the optical spectrum and the PIN-based RF spectrum. We use an RF amplifier (12-GHz bandwidth) to amplify the photocurrent from the PIN, and the measured DC component of the 3PA-induced current is used to monitor the intracavity power. As the pump is tuned into resonance, we observe the formation of primary FWM sidebands [Fig. \ref{fig:fig4}(a)(i)]. With further pump detuning, we observe the formation and interaction of multiple mini-combs, resulting in an increase in RF amplitude noise [Fig. \ref{fig:fig4}(a)(ii) \& (iii)]. As the pump is tuned further, an abrupt transition in the optical spectrum occurs along with a reduction of the RF amplitude noise [Fig. \ref{fig:fig4}(a)(iv)], which is consistent with previous observations of phase-locking in other platforms \cite{Saha,Herr}. Additionally, Fig. \ref{fig:fig4}(b) shows the DC component of the 3PA-induced current for various pump detunings. As the comb evolves to the high-noise state [Fig. \ref{fig:fig4}(a)(i)--(iii)], we see a steady increase in DC current. When the comb undergoes a transition to the low-noise state, we observe an abrupt increase in DC current from 0.945 to 1.17 mA, which is suggestive of modelocking and pulse formation since the 3PA-induced DC current is very sensitive to temporal peak power in the cavity. To our knowledge, this is the first evidence of a coherent, low noise microresonator-based frequency comb demonstrated in the mid-IR. 

\begin{figure}[tp]
\centering\includegraphics[width=12cm]{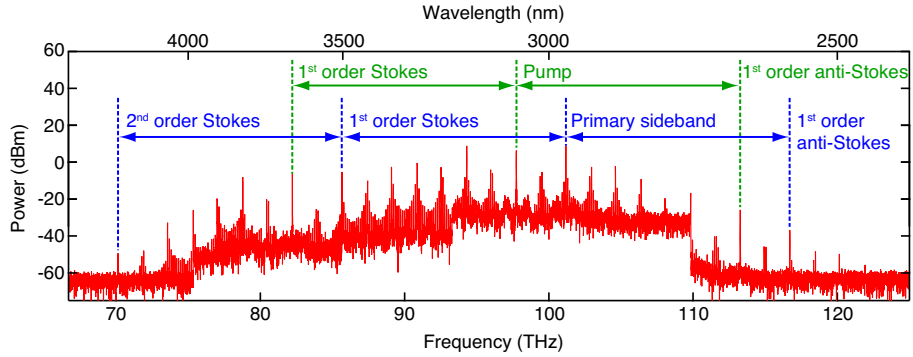}
\caption{Optical spectrum of coherent mid-IR comb generation in a silicon microresonator. The generated comb shows interplay between FWM and SRS. The Raman interaction with respect to the pump mode is shown in green. The interaction with respect to the primary sideband is shown in blue.
}
\label{fig:fig5}
\end{figure}

Figure 5 shows the low-noise comb spectrum in the frequency domain. We achieve a nearly-octave-spanning frequency comb spanning 70--122 THz, which corresponds to 2460--4278 nm in wavelength. Interestingly, the transition to the low-noise state coincides with the emergence of multiple Raman oscillations, implying that the low-noise state is strongly influenced by simulated Raman scattering (SRS), The Raman effect has been previously observed and characterized in silicon nanowaveguides with a Raman frequency shift of 15.6 THz \cite{RongLaser05,Claps} and a narrow linewidth of 105 GHz. Previous theoretical investigations have indicated that comb generation based on the Raman effect is possible with the comb spacing defined by the Raman shift \cite{Hansson}. In our experiment, we observe distinct Stokes and anti-Stokes peaks separated by 15.52 THz with respect to the pump frequency (indicated with green arrows in Fig. \ref{fig:fig5}). Furthermore, other strong comb lines are separated by the Raman shift, and we observe cascaded Raman Stokes lines with respect to the primary comb line (indicated with blue arrows in Fig. \ref{fig:fig5}). Generation of higher-order Raman lines is limited by the high optical losses due to the silica cladding beyond 4.3 \textmu m. Another indication of the SRS process is the significantly depleted pump frequency line with power lower than the primary FWM sideband. The fact that the frequency shift of the observed Raman peaks is lower than the value of 15.6 THz at room temperature could be attributed to: 1) the spectral overlap between the cavity resonance and the Raman gain profile 2) and the decreased Raman shift in silicon with higher temperatures \cite{Hart}, which is due to absorption and 3PA-induced thermal effects in the silicon microresonator. The large FSR of our device (127 GHz), along with the large Raman shift and narrow Raman gain bandwidth, makes the generation of Raman-assisted frequency comb much more sensitive to the pump laser detuning and the pump power. Another feature of this Raman-assisted frequency comb is that, in contrast to Hansson, \emph{et al}. \cite{Hansson}, our generated comb shows discrete high power comb lines with a spacing of 1.7 THz, which corresponds to 13 FSR's and 1/9 of the Raman frequency shift. The spacing is defined by the interplay of FWM and SRS when pumping at anomalous GVD. The high power comb lines that emerge in this final state dominates the phase locking and pulse formation. Our results indicate that, in contrast to conventional FWM-induced phase-locked frequency combs, phase-locking in our system occurs as a result of a combination of coherent generation of Stokes and anti-Stokes frequencies from SRS \cite{YWang}, and FWM interactions between the different Stokes and anti-Stokes pairs, enabling broadband coherent frequency comb generation. 


In conclusion, we demonstrate near-octave spanning, coherent mid-IR frequency comb generation in a silicon microresonator. The phase-locking process is a result of both FWM and SRS resulting in a high repetition rate comb. The integrated PIN structure allows for direct probing of the RF characteristics of the comb generation dynamics by exploiting siliconÕs intrinsic 3PA loss. We believe these results represent a significant step toward a fully integrated frequency comb source in the mid-IR regime.

\section*{Acknowledgments} This work was performed in part at the Cornell Nanoscale Facility, a member of the National Nanotechnology Infrastructure Network, which is supported by the NSF (grant ECS-0335765). This material is based upon work supported by the Air Force Office of Scientific Research under award number FFA9550-15-1-0303. The authors also gratefully acknowledge support from the Intelligence Advanced Research Projects Activity (IARPA), Defense Advanced Research Projects Agency (W31P4Q-15-1-0015), and National Science Foundation (ECS-0335765, ECCS-1306035). 

\end{document}